\documentclass[11pt]{article}
\usepackage{fullpage}
\usepackage{amsmath}
\usepackage{amsfonts}
\usepackage[dvips]{epsfig}

\def\##1{\underline{#1}}
\def\=#1{\underline{\underline{#1}}}

\def\+#1{\underline{\bf #1}}
\def\*#1{\underline{\underline{\bf #1}}}

\def\r#1{(\ref{#1})}
\def\l#1{\label{#1}}
\def\c#1{\cite{#1}}

\def\le{\left(}
\def\ri{\right)}
\def\les{\left[}
\def\ris{\right]}
\def\lec{\left\{}
\def\ric{\right\}}

\def\.{\mbox{ \tiny{$^\bullet$} }}

\def\eps{\epsilon}

\def\I{\*I}

\begin{document}

\begin{center}
\Large{\bf {\LARGE On extended homogenization formalisms for
nanocomposites }}

\normalsize \vspace{6mm}

Tom G. Mackay\footnote{ Tel: +44 131 650 5058; fax: +44 131 650
6553; e--mail: T.Mackay@ed.ac.uk}

\vspace{4mm}

\noindent{ \emph{School of Mathematics,  University of Edinburgh,\\
James Clerk Maxwell Building, King's Buildings, Edinburgh EH9 3JZ,
United Kingdom.} }

\end{center} \vspace{12mm}

\begin{abstract}
In  a long wavelength regime, the effective properties of
particulate composites, including nanocomposites,  may be estimated
using one of various homogenization formalisms, such as the
Bruggeman and Maxwell Garnett formalisms, and the approach of the
strong--property--fluctuation theory (SPFT). In the conventional
implementations of these formalisms, the constituent particles are
treated as point--like scattering centres. However, extended
formalisms have been established~---~which involve integral
formulations~---~that take account of the spatial extent of the
constituent particles. In particular, the extended second--order
SPFT takes account of both the size of the constituent particles and
their statistical distributions. We derive explicit representations
of the extended second--order SPFT appropriate to isotropic chiral
and uniaxial dielectric homogenized composite mediums. These results
may also be employed in extended versions of the Bruggeman and
Maxwell Garnett formalisms.

\end{abstract}

{\bf keywords:} strong--property--fluctuation theory, depolarization
dyadic, Bruggeman formalism, Maxwell Garnett formalism

\section{Introduction}
\label{intro}  

In the present era of exotic composite materials with nanoscale
architectures, there is a pressing need for accurate theoretical
tools to predict their electromagnetic properties \c{Walser}. While
this generally poses a formidable challenge to theoreticians,
matters can be simplified considerably provided that wavelengths are
sufficiently long relative to the length scales of the particles
from which  the composite is assembled~---~the composite may then be
regarded as being effectively homogeneous. Various well established
formalisms are available to estimate the constitutive parameters of
such homogenized composite mediums (HCMs) \c{L96}. One of the most
sophisticated~---~and one which has also gained prominence lately in
studies of  HCMs as metamaterials \c{M05}~---~is based on the
strong--property--fluctuation theory (SPFT) \c{TK81}.

The origins of the SPFT lie in wave propagation studies pertaining
to continuous random mediums \c{Frisch,Ryzhov}, but it was later
 adapted to estimate the constitutive parameters of HCMs \c{TK81,Genchev,Z94,ML95,MLW00}.
Unlike other more commonly used approaches to homogenization, such
as  the Bruggeman and Maxwell Garnett formalisms \c{Ch6_WLM97}, the
SPFT can accommodate a  comprehensive description of the
distributional statistics of the constituent particles. The SPFT
provides an estimate of the HCM constitutive parameters via a
recursive scheme, based on an apt ambient medium described by the
Bruggeman formalism. Usually the recursive scheme is truncated at
the second--order level, wherein the spatial distribution statistics
of the constituent particles are described in terms of a two--point
covariance function and its associated correlation length. In a
recent development,  the \emph{extended} SPFT was established which
takes into account the spatial extent of the constituent particles
\c{M04,CM06}. Similarly extended versions of the Maxwell Garnett and
Bruggeman formalisms have also been developed
\c{SL93a,SL93b,Prinkey,S96,Mallet}, but these apply only to
isotropic HCMs (whereas the extended SPFT is available for
bianisotropic HCMs \c{CM06}) and involve simplistic descriptions of
the constituent particle distributions.

A drawback with the SPFT approach to homogenization is that its
implementation can be an involved process, generally requiring
numerical methods. Explicit expressions
 (i.e., ones not expressed in terms of integrals)
are not available for  SPFT
estimates of  HCM constitutive parameters, with the exception of
certain isotropic dielectric homogenization scenarios \c{TK81}. We
address this issue in the following by deriving explicit expressions
for the extended SPFT appropriate to isotropic chiral and uniaxial
dielectric HCMs. The expressions  derived for depolarization dyadics
can also be utilized in extended versions of the Bruggeman formalism
and the Maxwell Garnett formalism.

In the following,  vector quantities are underlined. Double
underlining  and normal (bold) face signify a 3$\times$3
(6$\times$6) dyadic.
 The determinant, inverse and transpose    of
a dyadic $\=M$ are denoted by $\mbox{det}\,\les \, \=M \, \ris $,
$\=M^{-1}$ and   $\=M^{T}$, respectively. The 3$\times$3
(6$\times$6) identity dyadic
 is represented by  $\,\=I\,$ ($\,\*I\,$); and
the 3$\times$3 (6$\times$6) null dyadic
 is represented by  $\,\=0\,$ ($\,\*0\,$).
Angular frequency is denoted by $\omega$; the permeability of free
space is $\mu_{\mbox{\tiny{0}}}$; and $i = \sqrt{-1}$.  The
homogeneous bianisotropic medium specified by the
(frequency--domain) Tellegen constitutive relations \c{ML_Prog_Opt}
\begin{equation}
\left.
\begin{array}{l}
\#D(\#r) = \=\eps_{\,\ell} \. \#E(\#r) + \=\xi_{\,\ell} \. \#H(\#r) \\
\#B(\#r) = \=\zeta_{\,\ell} \. \#E(\#r) + \=\mu_{\,\ell} \. \#H(\#r)
\end{array}
\right\}
\end{equation}
is compactly characterized  by  its 6$\times$6 constitutive dyadic
\begin{equation}
\*K_{\,\ell} = \les \begin{array}{cc} \=\eps_{\,\ell} &
\=\xi_{\,\ell}
\\ \=\zeta_{\,\ell} & \=\mu_{\,\ell} \end{array} \ris,
\end{equation}
which subsumes the four 3$\times$3 constitutive dyadics
$\=\eps_{\,\ell}$, $\=\xi_{\,\ell}$, $\=\zeta_{\,\ell}$ and
$\=\mu_{\,\ell}$. Subscripts   are used to identify the particular
medium that the constitutive dyadics describe.

\section{Analysis}
\label{analysis}  

Many approaches to homogenization, including those of the SPFT and
the Bruggeman and Maxwell Garnett formalisms, rely on depolarization
dyadics to  represent the scattering responses of the constituent
particles.  General integral formulations of depolarization dyadics
are presented in  \S\ref{depol_section}; and we show how these are
incorporated into the SPFT in \S\ref{Homog_section}. In
\S\ref{explicit_section} the main results of this communication are
presented as explicit representations of the extended SPFT for
isotropic chiral HCMs and uniaxial dielectric HCMs.

\subsection{Depolarization region}

\l{depol_section}

Let us consider a homogeneous spherical particle of radius $\eta$,
embedded in a homogeneous ambient medium characterized by the
6$\times$6 constitutive dyadic $\*K_{\,\mbox{\tiny{amb}}}$. Provided
that:
\begin{itemize}
\item[(i)]  current density distributions
induced within the particle are uniform throughout its volume, and
\item[(ii)]
the particle is   small relative to electromagnetic wavelengths,
\end{itemize}
the particle's scattering response is captured by the depolarization
dyadic \c{M97,WSW99}
\begin{equation}
\*D (\eta) = \int_{|\#r| < \eta} \, \*G_{\,\mbox{\tiny{amb}}} (\#r)
\; d^3 \#r . \l{depol_def}
\end{equation}
Herein,  $\*G_{\,\mbox{\tiny{amb}}} (\#r)$ is the 6$\times$6  dyadic
Green function of the ambient medium. While explicit representations
of $\*G_{\,\mbox{\tiny{amb}}} (\#r)$ are  not generally available
for anisotropic and bianisotropic ambient mediums \c{W93}, its
Fourier transform, namely
\begin{equation} \l{def_FT}
\*{\tilde{G}}_{\,\mbox{\tiny{amb}}}  (\#q) = \int_{\#r}
\*G_{\,\mbox{\tiny{amb}}}  (\#r) \, \exp (- i \#q \. \#r ) \; d^3
\#r \,,
\end{equation}
 is  expressible as \c{MW97}
\begin{equation}
\underline{\underline{\tilde{\bf G}}}_{\,\mbox{\tiny{amb}}}
(\#q) =
\frac{1}{i \omega} \, \les \underline{\underline{\tilde{\bf
A}}}_{\,\mbox{\tiny{amb}}} (\#q) \ris^{-1},
\end{equation}
where
\begin{equation}
\underline{\underline{\tilde{\bf A}}}_{\,\mbox{\tiny{amb}}} (\#q)=
\les
\begin{array}{cc} \=0 &  (\#q /\omega) \times \=I \\ \vspace{-8pt} & \\
-(\#q /\omega) \times \=I & \=0
\end{array} \ris +
\underline{\underline{\bf K}}_{\,\mbox{\tiny{amb}}}. \l{Gq_Ad}
\end{equation}
Exploiting the spectral representation \r{def_FT}, the
depolarization dyadic is given by
 \c{M97, MW97}
\begin{eqnarray}
\*D (\eta) &=&
 \frac{\eta}{2 \pi^2} \, \int_{\#q} \frac{1}{q^2} \, \les \frac{\sin
(q \eta )}{q \eta} - \cos ( q \eta) \ris \,
\*{\tilde{G}}_{\,\mbox{\tiny{amb}}} (\#q) \; d^3 \#q ,
\end{eqnarray}
with $q^2 = \#q \. \#q$.

In order to accommodate  a  depolarization region of nonzero volume,
the depolarization dyadic may be considered as the sum \c{M04,CM06}
\begin{equation}
\*D (\eta) =    \*D^{ 0} + \*D^{+} (\eta) , \l{D_eta_def}
\end{equation}
wherein
\begin{eqnarray}
&& \*D^{ 0} =  \frac{\eta}{2 \pi^2} \, \int_{\#q} \frac{1}{q^2} \,
\les \frac{\sin (q \eta )}{q \eta} - \cos ( q \eta) \ris \,
\*{\tilde{G}}^{ \infty}_{\,\mbox{\tiny{amb}}} (\hat{\#q}) \; d^3 \#q
\, ,
\l{Dinf}\\
&& \*D^{+} (\eta) =  \frac{\eta}{2 \pi^2} \, \int_{\#q}
\frac{1}{q^2} \, \les \frac{\sin (q \eta )}{q \eta} - \cos ( q \eta)
\ris \, \*{\tilde{G}}^{+}_{\,\mbox{\tiny{amb}}} (\#q) \; d^3 \#q \,
, \l{D0}
\end{eqnarray}
with
\begin{eqnarray}
\underline{\underline{\tilde{\bf G}}}^{\infty}_{\,\mbox{\tiny{amb}}}
(\hat{\#q}) &=& \lim_{q\rightarrow \infty} \;
\underline{\underline{\tilde{\bf G}}}_{\,\mbox{\tiny{amb}}} (\#q),
\\
 \underline{\underline{\tilde{\bf G}}}^{+}_{\,\mbox{\tiny{amb}}}
 (\#q)
&=& \underline{\underline{\tilde{\bf G}}}_{\,\mbox{\tiny{amb}}}
(\#q) - \underline{\underline{\tilde{\bf
G}}}^{\infty}_{\,\mbox{\tiny{amb}}} (\hat{\#q}), \l{G_Go_Gi}
\end{eqnarray}
and the unit vector $\hat{\#q} = \#q / q$$ = ( \sin \theta \cos
\phi,$  $\sin \theta \sin \phi, $  $ \cos \theta )$.
 The dyadic  $ \*D^{0}$ represents the depolarization
contribution arising from the vanishingly small spherical region in
the limit $ \eta \to 0 $, whereas the dyadic $ \*D^{+}(\eta)$
provides the depolarization contribution arising from the spherical
region of nonzero volume. It is widespread practice in
homogenization studies to neglect
 $ \*D^{+} (\eta)$ and simply
take $ \*D^{0}$ as the depolarization dyadic  \c{Michel00}. However,
the importance of
 the  spatial extent of depolarization
regions has been underlined in  studies of isotropic \c{Doyle,
Dungey,
 SL93a,Prinkey, S96}, anisotropic \c{M04} and bianisotropic \c{CM06} HCMs.

The integral representation of $ \*D^{0}$ has been extensively
studied \c{Michel00}. The volume integral in eq. \r{Dinf} simplifies
to the $\eta$--independent surface integral
  \c{M97,MW97}
\begin{equation} \l{dd_def}
\*D^0 = \frac{1}{4 \pi}\,  \int^{2 \pi}_{\phi = 0} \,
\int^{\pi}_{\theta = 0} \,\underline{\underline{\tilde{\bf
G}}}^{\infty}_{\,\mbox{\tiny{amb}}} (\hat{\#q})\;\; \sin \theta \;
d\theta \; d\phi ,
\end{equation}
which is  easily evaluated for  isotropic dielectric--magnetic and
isotropic chiral mediums \c{ML_Prog_Opt}. Explicit evaluations for
uniaxial dielectric  mediums have  been presented in terms of
hyperbolic functions \c{M97}, whereas an elliptic function
representation is available for biaxial dielectric mediums \c{W98}.

The depolarization contribution given by $ \*D^{+}(\eta)$ has been
computed in various numerical studies \c{M04,CM06}, but hitherto no
explicit evaluations of the volume integral
  in eq. \r{D0} have been reported for ambient mediums other than isotropic dielectric
ambient mediums \c{TK81}.
  A notable simplification arises in the case of Lorentz--reciprocal
ambient mediums (i.e.,  ambient mediums which satisfy
$\=\eps_{\,\mbox{\tiny{amb}}} = \=\eps^T_{\,\mbox{\tiny{amb}}}$, $
\=\xi_{\,\mbox{\tiny{amb}}} = - \=\zeta^T_{\,\mbox{\tiny{amb}}}$,
and $\=\mu_{\,\mbox{\tiny{amb}}} = \=\mu^T_{\,\mbox{\tiny{amb}}}$
\c{Krowne}). Therein  $\*D^{+}(\eta)$ has the
 surface integral representation \c{MLW00}
\begin{eqnarray} \l{D_plus}
\*D^{+}(\eta) &=& \frac{\omega^4}{4 \pi } \int^{2 \pi}_{\phi = 0}
\int^{\pi}_{\theta = 0} \frac{1}{b_{\,\mbox{\tiny{amb}}}
(\theta,
\phi) }\Bigg[ \frac{1}{ \kappa_+ - \kappa_- } \Bigg( \frac{e^{i \eta
q}}{2 q^2} \le 1 - i \eta q\ri \Big\{ \,
 \mbox{det} \les \underline{\underline{\tilde{\bf
A}}}_{\,\mbox{\tiny{amb}}}(\#q) \ris
\underline{\underline{\tilde{\bf G}}}^+_{\,\mbox{\tiny{amb}}}(\#q)
 \nonumber
\\ && +  \mbox{det} \les \underline{\underline{\tilde{\bf
A}}}_{\,\mbox{\tiny{amb}}}(-\#q) \ris
\underline{\underline{\tilde{\bf G}}}^+_{\,\mbox{\tiny{amb}}}(-\#q)
\Big\} \Bigg)^{q= \sqrt{\kappa_+}}_{q=\sqrt{\kappa_-}} + \frac{
 \mbox{det} \les \underline{\underline{\tilde{\bf
A}}}_{\,\mbox{\tiny{amb}}}(\#0) \ris} {\kappa_+ \kappa_-}\,
\underline{\underline{\tilde{\bf G}}}^+_{\,\mbox{\tiny{amb}}}(\#0)
 \Bigg]\, \sin
\theta \; d \theta \; d \phi, \nonumber \\ &&  \l{D_e}
\end{eqnarray}
with $\kappa_\pm$ being the $q^2$ roots of $\mbox{det} \les
\underline{\underline{\tilde{\bf A}}}_{\,\mbox{\tiny{amb}}}(\#q)
\ris = 0$ and the scalar function
\begin{equation}
b_{\,\mbox{\tiny{amb}}}
( \theta, \phi ) = \le \, \hat{\#q}\.
\={\eps}_{\,\mbox{\tiny{amb}}}  \.\hat{\#q}\, \ri \le \, \hat{\#q}\.
 \={\mu}_{\,\mbox{\tiny{amb}}}  \.\hat{\#q} \,
\ri + \le \, \hat{\#q}\. \={\xi}_{\,\mbox{\tiny{amb}}}  \.
\hat{\#q}\,\ri^2.
\end{equation}
Evaluations of the surface integral on the right side of eq. \r{D_e}
are provided in \S\ref{icm_section} and \S\ref{uni_section} for
isotropic chiral ambient mediums and uniaxial dielectric ambient
mediums, respectively.

\subsection{Homogenization }
\l{Homog_section}

Depolarization dyadics play a central role in formalisms employed
 to estimate the constitutive parameters of HCMs \c{TK81}.
Here we focus on the approach provided by the extended SPFT
\c{M04,CM06}.
  We concentrate on the homogenization of a two--phase composite
 wherein the two constituent phases, labelled as $a$ and $b$,
comprise  spherical particles of average radius $\eta $. The
constituent phase $a$ occupies the region $V_{\mbox{\tiny{a}}}$
whereas constituent phase $b$ occupies the region
$V_{\mbox{\tiny{b}}}$, The constituent phases
 are randomly mixed with their
 distributional statistics being described in
terms of moments of the characteristic functions
\begin{equation}
\Phi_{ \ell}(\#r) = \left\{ \begin{array}{ll} 1, & \qquad \#r \in
V_{\, \ell},\\ & \qquad \qquad \qquad \qquad \qquad \qquad (\ell=\mbox{a,b}) . \\
 0, & \qquad \#r \not\in V_{\, \ell}, \end{array} \right.
\end{equation}
 The volume fraction of phase $\ell$, namely $f_\ell$ , is given by
the first statistical moment of
 $\Phi_{\ell}$ ;
 i.e., $\langle \, \Phi_{\ell}(\#r) \, \rangle = f_\ell$ . We have
 $\langle \, \Phi_{\mbox{\tiny{a}}}(\#r) \, \rangle + \langle \, \Phi_{\mbox{\tiny{b}}}(\#r) \, \rangle = 1$.
For the second statistical moment of $\Phi_{\ell}$, the
physically--motivated covariance form \c{TKN82}
\begin{equation}
\langle \, \Phi_\ell (\#r) \, \Phi_\ell (\#r')\,\rangle = \left\{
\begin{array}{lll}
\langle \, \Phi_\ell (\#r) \, \rangle \langle \Phi_\ell
(\#r')\,\rangle\,, & & \hspace{10mm}
  |     \#r - \#r'  | > L ,\\ && \hspace{45mm} (\ell=\mbox{a,b}), \\
\langle \, \Phi_\ell (\#r) \, \rangle \,, && \hspace{10mm}
 |   \#r -
\#r'  | \leq L ,
\end{array}
\right.
 \l{cov}
\end{equation}
is commonly implemented. Herein, the correlation length $L$
 is required to be  much smaller than the electromagnetic
 wavelengths, but much larger than the constituent particle
radius  $\eta$. Parenthetically,  the specific form of the
covariance function has only a secondary influence on SPFT estimates
of HCM constitutive parameters, across a range of
physically--plausible covariance functions \c{MLW01b}.

The $n$th--order SPFT estimate of the HCM constitutive dyadic,
namely $ \*K^{[n]}_{\,\mbox{\tiny{HCM}}}$,
 derives from the
recursive refinement of an ambient homogeneous  medium,
characterized by the constitutive dyadic
$\*K_{\,\mbox{\tiny{amb}}}$. At lowest order (i.e.,  zeroth  and
first order), the SPFT  estimate of the HCM constitutive dyadic is
identical to that of the ambient medium \c{MLW00}; i.e.,
\begin{equation}
\*K^{[0]}_{\,\mbox{\tiny{HCM}}} = \*K^{[1]}_{\,\mbox{\tiny{HCM}}}
 = \*K_{\,\mbox{\tiny{amb}}}.
\end{equation}
Furthermore,  $\*K_{\,\mbox{\tiny{amb}}}$ is delivered by solving
the nonlinear equations
\begin{equation}
f_{\mbox{\tiny{a}}} \,\mbox{\boldmath$\=\chi$}_{\,\mbox{\tiny{a}}}
(\eta)  + f_{\mbox{\tiny{b}}}
\,\mbox{\boldmath$\=\chi$}_{\,\mbox{\tiny{b}}} (\eta)  = \*0\,,
\l{Br}
\end{equation}
 wherein the polarizability density dyadics
\begin{equation} \l{polarizability}
 \mbox{\boldmath$\=\chi$}_{\, \ell} ( \eta)  = - i \omega
\le\,\*K_{\,\ell}  - \*K_{\,\mbox{\tiny{amb}}}  \,\ri\.\les \,\*I +
i \omega \*D (\eta) \. \le\, \*K_{\,\ell}  -
\*K_{\,\mbox{\tiny{amb}}} \,\ri \ris^{-1},
 \qquad  (\ell = \mbox{a,b}),
\end{equation}
and $\*K_{\,\ell}$  ($\ell = \mbox{a,b}$) are the constitutive
dyadics of constituent phases $a$ and $b$.

For a broad range of  bianisotropic HCMs, the SPFT (including the
extended SPFT \c{Cui_Mackay_3rd}) converges at the second-order
level\footnote{The second--order SPFT approximation is  also known
as the bilocal approximation.} \c{MLW01}. The second--order SPFT
estimate of the HCM constitutive dyadic is \c{MLW00}
\begin{equation}
 \*K^{[2]}_{\,\mbox{\tiny{HCM}}}  =
 \*K_{\,\mbox{\tiny{amb}}}  - \frac{1}{i \omega} \les \,\I +
\mbox{\boldmath$\={\Sigma}$}^{[2]} (\eta, L)  \. \*D (\eta)
\,\ris^{-1} \. \mbox{\boldmath$\={\Sigma}$}^{[2]} (\eta, L)
,\l{KDy0}
\end{equation}
with
 the mass operator dyadic term \c{Frisch}~---~corresponding to the covariance
 function \r{cov}~---~being
\begin{equation}
\underline{\underline{\mbox{\boldmath$\Sigma$}}}^{[2]} (\eta, L)   =
f_{\mbox{\tiny{a}}} f_{\mbox{\tiny{b}}} \les \,
\mbox{\boldmath$\=\chi$}_{\,\mbox{\tiny{a}}} (\eta) -
\mbox{\boldmath$\=\chi$}_{\,\mbox{\tiny{b}}} (\eta)
 \ris \.   \*D^{+} (L)
 \. \les \, \mbox{\boldmath$\=\chi$}_{\,\mbox{\tiny{a}}} (\eta) -
\mbox{\boldmath$\=\chi$}_{\,\mbox{\tiny{b}}} (\eta)   \ris. \l{spv}
\end{equation}
Thus, within the extended  second--order SPFT,  the estimate of the
HCM constitutive dyadic depends on two length scales: the
constituent particle size $\eta$ via the depolarization dyadic and
the correlation length $L$ via the mass operator.

\subsection{Explicit representations} \l{explicit_section}

We now turn to the evaluation of the surface integral which delivers
$\*D^+$. This term crops up as $\*D^+ (\eta)$ in the extended
depolarization dyadic representation \r{D_eta_def} and as $\*D^+
(L)$
 in the mass operator term \r{spv} which yields the second--order
 SPFT estimate of the HCM constitutive dyadic. A key step in
the evaluation of the integral on the right side of eq. \r{D_plus}
is the  approximation of the $e^{i \eta q} \le 1 - i \eta q\ri $
term in the integrand by its asymptotic expansion $1 + (\eta q)^2/2
+ i (\eta q)^3/3$, which is permissible since $\eta
\sqrt{\kappa_\pm} \ll 1$ in the long wavelength regime. The
evaluation of $\*D^+ (L)$ is isomorphic to that of  $\*D^+ (\eta)$
since here we similarly have $L \sqrt{\kappa_\pm} \ll 1$ in the long
wavelength regime.
 In the following
we consider two types of ambient medium: an isotropic chiral ambient
medium in \S\ref{icm_section} and a uniaxial dielectric ambient
medium in \S\ref{uni_section}.

\subsubsection{Isotropic chiral ambient medium} \l{icm_section}

Suppose that the ambient medium is an isotropic chiral medium.  Its
constitutive dyadic  has the form \c{Beltrami}
\begin{equation}
\*K_{\,\mbox{\tiny{amb}}} = \les
\begin{array}{cc}
\eps_{\mbox{\tiny{amb}}} \, \=I & \xi_{\mbox{\tiny{amb}}}\, \=I \vspace{4pt} \\
-\xi_{\mbox{\tiny{amb}}} \,\=I & \mu_{\mbox{\tiny{amb}}} \, \=I
\end{array}
\ris.
\end{equation}
The integral on the right side of eq. \r{D_plus} may then be
evaluated to give
\begin{eqnarray} \l{Dc}
\*D^+ (\eta) &=& \frac{i \omega \eta^2}{3 } \Bigg(
 \les
\begin{array}{cc}
 \mu_{\mbox{\tiny{amb}}}  \, \=I
 & \xi_{\mbox{\tiny{amb}}}\, \=I \vspace{4pt} \\
-\xi_{\mbox{\tiny{amb}}} \,\=I & \eps_{\mbox{\tiny{amb}}} \, \=I
\end{array}
\ris \nonumber \\ && + i \frac{2 \omega \eta}{3} \les
\begin{array}{cc}
\displaystyle{
\sqrt{\frac{\mu_{\mbox{\tiny{amb}}}}{\eps_{\mbox{\tiny{amb}}}}} \le
\eps_{\mbox{\tiny{amb}}} \, \mu_{\mbox{\tiny{amb}}} -
\xi^2_{\mbox{\tiny{amb}}} \ri  \, \=I}
 & 2 \xi_{\mbox{\tiny{amb}}} \sqrt{\eps_{\mbox{\tiny{amb}}} \,\mu_{\mbox{\tiny{amb}}} }\, \=I \vspace{4pt} \\
- 2 \xi_{\mbox{\tiny{amb}}} \sqrt{\eps_{\mbox{\tiny{amb}}} \,
\mu_{\mbox{\tiny{amb}}} }\,\=I & \displaystyle{
\sqrt{\frac{\eps_{\mbox{\tiny{amb}}}}{\mu_{\mbox{\tiny{amb}}}}} \le
\eps_{\mbox{\tiny{amb}}}\, \mu_{\mbox{\tiny{amb}}} -
\xi^2_{\mbox{\tiny{amb}}} \ri  \, \=I}
\end{array}
\ris \Bigg), \quad
\end{eqnarray}
after some straightforward manipulations.

\subsubsection{Uniaxial dielectric ambient medium} \l{uni_section}

Suppose that the ambient medium is a uniaxial dielectric medium.
Without loss of generality, let us take the distinguished axis of
this uniaxial medium to be aligned with the Cartesian $z$ axis.
Accordingly,  its constitutive dyadic  takes the form
\c{ML_Prog_Opt}
\begin{equation}
\*K_{\,\mbox{\tiny{amb}}} = \les
\begin{array}{cc}
\le \begin{array}{ccc} \eps^{}_{\mbox{\tiny{amb}}} &0 &0 \\
0& \eps^{}_{\mbox{\tiny{amb}}} & 0 \\
0& 0& \eps^z_{\mbox{\tiny{amb}}} \end{array} \ri &  \=0 \vspace{4pt} \\
\=0 & \mu_{\mbox{\tiny{0}}} \, \=I
\end{array}
\ris.
\end{equation}
Within the present context of the SPFT, such an ambient medium
arises in homogenization scenarios in which one of the constituent
phases is a uniaxial dielectric medium and the other is either an
isotropic dielectric medium or a uniaxial dielectric medium. Hence,
we observe from eqs. \r{polarizability} and \r{spv} that only the
upper three diagonal entries of $\*D^+$, namely $\les \, \*D^+
\ris_{nn}$ ($n=1,2,3$), contribute to the estimate of the HCM
constitutive dyadic. After some manipulations, these entries are
evaluated as
\begin{eqnarray}
\les \, \*D^+ (\eta) \ris_{11} &=& \les \, \*D^+ (\eta) \ris_{22}
\nonumber \\ &=&  \frac{i \omega \mu_{\mbox{\tiny{0}}} \eta^2}{8}
\Bigg[ \frac{1 -
\gamma_{\mbox{\tiny{amb}}}}{\gamma_{\mbox{\tiny{amb}}}} - \le
\frac{\eps^z_{\mbox{\tiny{amb}}}}{\gamma_{\mbox{\tiny{amb}}}
\eps^{}_{\mbox{\tiny{amb}}}} \ri^2 \sqrt{
\gamma_{\mbox{\tiny{amb}}}}\, \tanh^{-1} \le
\sqrt{\gamma_{\mbox{\tiny{amb}}}} \ri \nonumber \\ && + i \eta
\frac{4 \omega \le 3 \eps^{}_{\mbox{\tiny{amb}}} +
\eps^z_{\mbox{\tiny{amb}}} \ri}{9}
\sqrt{\frac{\mu_{\mbox{\tiny{0}}}}{\eps^{}_{\mbox{\tiny{amb}}}}} \Bigg], \l{Dx} \\
 \les \, \*D^+ (\eta) \ris_{33} &=& \frac{i \omega \mu_{\mbox{\tiny{0}}} \eta^2}{4}
  \lec \frac{1}{\gamma_{\mbox{\tiny{amb}}}} \les \frac{1 +
\gamma_{\mbox{\tiny{amb}}}}{\gamma_{\mbox{\tiny{amb}}}} \sqrt{
\gamma_{\mbox{\tiny{amb}}}}\, \, \tanh^{-1} \le
\sqrt{\gamma_{\mbox{\tiny{amb}}}} \ri -1 \ris + i \eta \frac{ 4
\omega \sqrt{
\eps^{}_{\mbox{\tiny{amb}}}\,\mu_{\mbox{\tiny{0}}}}}{9} \ric ,
\nonumber \\  \l{Dz} &&
\end{eqnarray}
with the dimensionless scalar
\begin{equation}
\gamma_{\mbox{\tiny{amb}}} = \frac{\eps^{}_{\mbox{\tiny{amb}}} -
\eps^z_{\mbox{\tiny{amb}}}}{\eps^{}_{\mbox{\tiny{amb}}}}.
\end{equation}
The representations \r{Dx} and  \r{Dz} apply when
$\gamma_{\mbox{\tiny{amb}}} $ is complex--valued  (with nonzero
imaginary part). If $\gamma_{\mbox{\tiny{amb}}}$ is real--valued
then the representations \r{Dx} and  \r{Dz} apply when $0 <
\gamma_{\mbox{\tiny{amb}}} < 1$; for $\gamma_{\mbox{\tiny{amb}}} <
0$, the $\sqrt{ \gamma_{\mbox{\tiny{amb}}}}\, \, \tanh^{-1} \le
\sqrt{\gamma_{\mbox{\tiny{amb}}}} \ri$ term in eqs. \r{Dx} and
\r{Dz} should be replaced by $- \sqrt{-
\gamma_{\mbox{\tiny{amb}}}}\, \, \tan^{-1} \le
\sqrt{-\gamma_{\mbox{\tiny{amb}}}} \ri$. In the scenario $
\gamma_{\mbox{\tiny{amb}}}>1$~---~which corresponds to
nondissipative uniaxial dielectric ambient mediums with indefinite
permittivity dyadics \c{MLD_06,Li_Ch4,Smith03}~---~the components of
$\*D^+ (\eta)$ are undefined, as are the corresponding components of
$\*D^0 $ \c{M97}.

\section{Discussion}
\l{discussion_section}

The main results of this communication are the derivations of the
eqs. \r{Dc}, \r{Dx} and \r{Dz} which, when substituted into eq.
\r{KDy0}, yield explicit formulations of the extended second--order
SPFT for isotropic chiral HCMs and uniaxial dielectric HCMs (when
supplemented with the corresponding expressions for $\*D^0$ which
are available elsewhere \c{ML_Prog_Opt}). For more complex HCMs,
numerical methods are needed to evaluate the depolarization dyadic
$\*D (\eta)$ and mass operator
$\underline{\underline{\mbox{\boldmath$\Sigma$}}}^{[2]} (\eta, L)$.

It is helpful to consider the isotropic dielectric specialization of
the extended second--order  SPFT result \r{KDy0}. In this  case, the
constituent phases are both isotropic dielectric mediums with
permittivities $\eps^{}_{\mbox{\tiny{a}}}$ and
$\eps^{}_{\mbox{\tiny{b}}}$; in consonance,  the ambient medium is
also an isotropic dielectric medium with permittivity
$\eps^{}_{\mbox{\tiny{amb}}}$.
 The
corresponding second--order SPFT estimate of the HCM permittivity is
\begin{equation}
\eps^{[2]}_{\mbox{\tiny{HCM}}}= \eps^{}_{\mbox{\tiny{amb}}} -
\frac{1}{i \omega} \le \frac{\Sigma^{[2]} (\eta, L)}{1 +
\Sigma^{[2]} (\eta, L) \, d(\eta)} \ri, \l{iso_1}
\end{equation}
wherein  the depolarization scalar may be expressed as the sum
\begin{equation}
d(\eta) = d^0 + d^+ (\eta).
\end{equation}
 The  contribution to $d (\eta)$
 associated with a vanishingly small
 depolarization region is provided by the well known result \c{ML_Prog_Opt}
\begin{equation}
d^0 = \frac{1}{i 3  \omega  \eps^{}_{\mbox{\tiny{amb}}}},
\end{equation}
whereas the contribution associated  with a depolarization region of
nonzero volume  may be extracted from eqs. \r{Dx} and \r{Dz}
 in the limit
$\eps^z_{\mbox{\tiny{amb}}} \to \eps^{}_{\mbox{\tiny{amb}}}$ as
\begin{equation}
 d^+ (\eta)
=   \frac{i   \mu_{\mbox{\tiny{0}}} \omega \eta^2 }{9} \le 3 + i 2
\eta \, \omega \sqrt{\eps^{}_{\mbox{\tiny{amb}}} \,
\mu_{\mbox{\tiny{0}}}} \, \ri.
\end{equation}
This expression for $d^+ (\eta)$  also follows from eq. \r{Dc} when
$\zeta_{\mbox{\tiny{amb}}} = 0$ and
 $\mu_{\mbox{\tiny{amb}}} =
\mu_{\mbox{\tiny{0}}}$. The scalar mass operator term in eq.
\r{iso_1} is provided by the isotropic dielectric specialization of
eq. \r{spv} as
\begin{equation}
\Sigma^{[2]} (\eta, L) = f_{\mbox{\tiny{a}}} f_{\mbox{\tiny{b}}}
\les \chi_{\mbox{\tiny{a}}} (\eta) - \chi_{\mbox{\tiny{b}}} (\eta)
\ris^2 d^+ (L),
\end{equation}
with the polarizability density scalars
\begin{equation} \l{polarizability2}
 \chi_{\ell} ( \eta)  = - i \omega
\les \frac{\eps_{\ell}  - \eps_{\mbox{\tiny{amb}}}}{1 + i \omega
\le \eps_{\ell}  - \eps_{\mbox{\tiny{amb}}} \ri \, d (\eta)} \ris
,
 \qquad  (\ell = \mbox{a,b}).
\end{equation}
We note that the explicit expression for the  extended second--order
SPFT estimate of the HCM permittivity \r{iso_1} is consistent with a
corresponding  result derived for the \emph{unextended} SPFT, within
the long wavelength limit represented by \c{TK81}
\begin{equation}
\eps^{[2]}_{\mbox{\tiny{HCM}}}= \eps^{}_{\mbox{\tiny{amb}}} -
\frac{1}{i \omega} \Sigma^{[2]} (L), \l{iso_2}
\end{equation}
where $\Sigma^{[2]} (L) \equiv \Sigma^{[2]} (\eta, L)$ evaluated
with $d^0$ in lieu of $d (\eta)$.

While our focus has primarily been on the extended second--order
SPFT, we  bear in mind that the expressions \r{Dc}, \r{Dx} and
\r{Dz} can also be deployed in extended versions of
  other  homogenization
formalisms, such as the frequently used Bruggeman and Maxwell
Garnett formalisms.
  In fact, the Bruggeman homogenization
formalism is equivalent to the zeroth order SPFT \c{ML95}. Hence the
extended Bruggeman estimate of the HCM constitutive dyadic is simply
the dyadic $\*K_{\,\mbox{\tiny{amb}}}$ which may be extracted from
eq. \r{Br}. Like the Bruggeman formalism, the Maxwell Garnett
formalism
is based on depolarization dyadics \c{Michel00}, but with
 one of the constituent phases playing the role of the ambient medium.
Therefore, the  Maxwell Garnett formalism~---~including its
incremental \c{Ch6_L98} and differential \c{Ch6_MLWM01}
variants~---~may be extended by implementing the appropriate
$\eta$--dependent depolarization dyadic $\*D (\eta)$.

We close by considering the question: For what range of  constituent
particle sizes are the extended homogenization formalisms discussed
herein appropriate? An upper bound is straightforwardly established
by the requirement that the particles must be small relative to the
electromagnetic wavelength(s), in order to be consistent with the
notion of homogenization. For optical wavelengths, the linear
dimensions of the constituent particles must therefore be at most
38--78 nm. A lower bound comes into effect because, at sufficiently
small length scales, quantum processes cannot be neglected in the
description of the constituent particles and their interactions.
Accordingly, the lower bound is material--dependent. We note that
for very small constituent particles, their constitutive parameters
may differ significantly from those of the corresponding bulk
materials and depend upon the shape and size of the particles
\c{Bohren}. This is particularly the case for metallic particles
smaller than the mean free path of conduction electrons in the bulk
metal, wherein the mean free path may be dominated by collisions at
the particle boundary \c{Kreibig}. Within this particle--size
regime, a recent study using spectroscopic ellipsometry demonstrated
that an extended Maxwell Garnett homogenization formalism adequately
characterizes a HCM based on silver nanoparticles as small as $2.3$
nm \c{Oates}. This study also highlights the prospects of
implementing extended homogenization formalisms, such as those
described herein, in particle sizing applications for
nanocomposites.


\end{document}